\def\bey{\begin{eqnarray}}
\def\eey{\end{eqnarray}}
\def\be{\begin{equation}}
\def\ee{\end{equation}}
\def\ba{\begin{array}}
\def\ea{\end{array}}
\def\gm{\gamma}

\def\Ld{\Lambda}
\def\af{\alpha}
\def\sg{\sigma}

\def\om{\omega}
\def\r{\rho}

\def\vep{\varepsilon}

\def\pp{\partial}
\def\pp{\partial}

\def\nnb{\nonumber}
\bigskip
\documentclass[onecolumn,showpacs,preprintnumbers,amsmath,amssymb]{revtex4}
\usepackage{graphicx}
\usepackage{fancyhdr}
\usepackage{dcolumn}
\usepackage{bm}
\usepackage{amssymb}
\usepackage{amsmath}
\usepackage{graphicx}
\usepackage[normalem]{ulem}
\usepackage[dvips]{color}

\begin{document}
\preprint{ }
\title{ Equation of state and hybrid star properties with the weakly interacting light U-boson in relativistic models}
\author{Dong-Rui Zhang$^1$, Wei-Zhou Jiang$^1$, Si-Na Wei$^1$, Rong-Yao Yang$^1$, Qian-Fei Xiang$^2$}
\affiliation{  $^1$ Department of Physics, Southeast University,
Nanjing 211189, China\\
$^2$Institute of High Energy Physics, Chinese Academy of Sciences, Beijing 100049, China
}

\begin{abstract}
\baselineskip18pt It has been a puzzle whether quarks may exist in
the interior of massive neutron stars, since the  hadron-quark
phase transition softens the equation of state (EOS) and  reduce
the neutron star (NS) maximum mass very significantly. In this
work, we consider the light U-boson that increases the NS maximum
mass appreciably through its weak coupling to fermions. The
inclusion of the U-boson may thus allow the existence of the quark
degrees of freedom in the interior of large mass neutron stars.
Unlike the consequence of the U-boson in hadronic matter,  the
stiffening role of the U-boson in the hybrid EOS is not sensitive
to the choice of the hadron phase models. In addition, we have
also investigated the effect of the effective QCD correction on
the hybrid EOS. This  correction may reduce the coupling strength
of the U-boson that is needed to satisfy NS maximum mass
constraint. While the inclusion of the U-boson also increases the
NS radius significantly, we find that appropriate in-medium
effects of the U-boson may reduce the NS radii significantly,
satisfying both the NS radius and mass constraints well.

\end{abstract}
\pacs{26.60.Kp, 21.60.Jz, 97.60.Jd} \keywords{Nuclear equation of state,
quark phase, U boson,  neutron stars} \maketitle \baselineskip
20.6pt

\section{Introduction}

The equation of state (EOS) of isospin asymmetric nuclear matter
is of prime importance for the investigation of nuclear
structure~\cite{Hor01,jia10,jia10a,wang14}, heavy-ion  reaction
dynamics~\cite{Li01,Li08}, and many issues in
astrophysics~\cite{Li08,lat00,lat01,lat04,lat07,Ste05}. However,
the EOS of asymmetric matter at supra-normal densities are still
poorly known~\cite{Li08,li15}, though the constraint on the
symmetric part of the EOS at supra-normal densities can be
extracted from the collective flow data of high energy heavy-ion
reactions~\cite{da02}. Different models and approaches can produce
rather different high-density
behaviors~\cite{br00,chen05,fu06,lie7}, while the complexity
arises for the EOS of asymmetric matter when the phase
transitions, such as hyperon productions,  meson condensations,
and quark deconfinement, take place at high densities. Normally,
phase transitions can   soften the EOS and reduce the neutron star
(NS) maximum mass significantly.

Recently, massive NS's, for instance, the pulsars PSR J1614-2230
with $M=1.97\pm0.04M_\odot$~\cite{Demorest2010}, and PSR
J0348+0432, with $M=2.01\pm0.04M_\odot$~\cite{Antoniadis2013} were
identified with the high-precision measurements. The larger NS
mass means the stiffer EOS of NS matter at high
densities. Considering the phase transitions in the NS core and
the EOS constraint from the collective flow data at high
densities, it is difficult for the nuclear models to reproduce
NS's as massive as $2M_\odot$. This indeed proposes a challenge to
the nuclear EOS at supra-normal densities. One may doubt whether
there are new degrees of freedom besides nucleons in
NS's~\cite{Demorest2010,Trumper04,Ozel06}. On the other hand, one
may contemplate what interactions can allow the new degrees of
freedom in massive NS's\cite{Bednarek11,jiang12,Ruster04,
Horvath04, Alford2005, Alford2007,Ippolito08, Fischer10,
Kurkela10a,Kurkela10b,Weissenborn11,Bonanno12,Yasutake14,Xia14}.

The mixture of  low-density hadronic matter and high-density quark
matter may form in the NS core after the hadron-quark transition
occurs with the  increase of density. In order to evade the
potential confliction between the softened EOS due to the
appearance of the quark phase and the observed massive NS's, it is
necessary to stiffen the quark EOS, for instance, by considering
the strong coupling and/or color superconductivity~\cite{Ruster04,
Horvath04, Alford2005, Alford2007,Ippolito08, Fischer10,
Weissenborn11}. Recently, a new repulsion, provided by the
U-boson, was introduced for nucleons in
NS's~\cite{Kr09,Wen09,zhang11}.  The light U-boson, first proposed
by Fayet~\cite{fa80}, might be regarded as the mediator of the
putative fifth force~\cite{fa86,fi99,ad03}. Recently, this light
U-boson has been considered as the interaction mediator of the MeV dark
matter  to account for the bright 511 keV $\gamma$-ray
from the galactic bulge~\cite{Bo04,Zhu07,Fa07,Je03}. The coupling
of the U-boson with the nucleons is very weak, but can increase
appreciably the NS maximum mass~\cite{Kr09,Wen09,zhang11} and
influence the transition density at the inner edge separating the
liquid core from the solid crust, effectively~\cite{Zheng12}.  In
particular, the weak coupling to baryons plays a striking role in
stabilizing neutron stars in the presence of the super-soft
symmetry energy~\cite{Wen09} that is extracted from  the FOPI/GSI
data on the $\pi^{-}/\pi^{+}$ ratio in relativistic heavy-ion
collisions~\cite{Xiao08}. 

In this work, we  consider the
effect of the U-boson on the softened EOS due to the hadron-quark
phase transition. We adopt the relativistic mean-field (RMF) theory, which achieved
great success in the past few
decades~\cite{Wal74,Bog77,Chin77,Ser86,Rei89,Ring96,Ser97,Ben03,Meng06,Ji07},
to describe hadronic matter,  and  the MIT bag model for quark
matter~\cite{ch74,he86}. For the hadron-quark phase transition, we use
Gibbs construction~\cite{Glendenning92,Glendenning01} to determine
the mixed phase of hadronic and quark matter. As the stiffening
role of the U-boson depends on the softness of the
models~\cite{zhang11}, we will examine how the U-boson stiffens
the hybrid EOS's initiated with different hadronic models. Since
quarks in the bag model are free of interaction,  it is  also interesting to
investigate  briefly the effect of  the effective correction from
the perturbative
QCD~\cite{Alford2005,Alcock86,Haensel86,Fraga01,xu15}. Eventually,
we will investigate the properties of hybrid stars with various
EOS's and discuss how the mass and radius constraints of the NS
observations can be satisfied. The paper is organized as follows.
In Sec.~\ref{rmf}, we present briefly the formalism. In
Sec.~\ref{results}, numerical results and discussions are
presented. At last, a summary is given in Sec.~\ref{summary}.

\section{Formalism}
\label{rmf}In the RMF models we adopted in this work, the
nucleon-nucleon interaction is realized via the exchange of three
mesons: the isoscalar meson $\sg$, which provides the intermediate-range
attraction between the nucleons, the isoscalar-vector meson $\om$,
which offers the short-range repulsion, and the isovector-vector
meson $b_0$, which accounts for the isospin dependence of the
nuclear force.  Though the $\pi$ meson interacts strongly with nucleons, we do not include it here because the
RMF framework just has the Hartree term that gives a zero contribution of the pseudoscalar meson.
The $\pi$ mesons can be included in the relativistic Hartree-Fock approximation. Without the exchange
terms, the RMF approximation still works well due mainly to the fact that in the
relativistic framework  the fermions are already identified by the
Dirac equation that is specifically for fermions. Moreover, the interaction in the RMF approximation, built upon
the equilibrium between the intermediate-range attraction and the short-range repulsion, can  yield the
saturation properties of nuclear matter very well~\cite{Wal74,Bog77,Chin77,Ser86,Rei89,Ring96,Ser97}.
The relativistic Lagrangian is then written as: \bey
 {\cal L}&=&
{\overline\psi}[i\gm_{\mu}\partial^{\mu}-M+g_{\sg}\sg-g_{\om}
\gm_{\mu}\om^{\mu}-g_\r\gm_\mu \tau_3 b_0^\mu
   ]\psi\nnb\\
      &  &
    - \frac{1}{4}F_{\mu\nu}F^{\mu\nu}+
      \frac{1}{2}m_{\om}^{2}\om_{\mu}\om^{\mu}
    - \frac{1}{4}B_{\mu\nu} B^{\mu\nu}
     +\frac{1}{2}m_{\r}^{2} b_{0\mu} b_0^{\mu}
    \\
     &&+
\frac{1}{2}(\partial_{\mu}\sg\partial^{\mu}\sg-m_{\sg}^{2}\sg^{2})
+U_{\rm eff}(\sg,\om^\mu, b_0^\mu)+{\cal L}_u\nnb, \label{eq:lag1}
  \eey
 where
 $\psi,\sg,\om$,$b_0$ are the fields of
the nucleon, scalar, vector, and neutral isovector-vector mesons,
with their masses $M, m_\sg,m_\om$, and $m_\r$, respectively.
$g_i(i=\sg,\om,\r)$  are the corresponding meson-nucleon
couplings. $F_{\mu\nu}$ and $ B_{\mu\nu}$ are the strength tensors
 of $\om$ and $\r$ mesons respectively,
\begin{equation}\label{strength} F_{\mu\nu}=\pp_\mu
\om_\nu -\pp_\nu \om_\mu,\hbox{  } B_{\mu\nu}=\pp_\mu b_{0\nu}
-\pp_\nu b_{0\mu}.
\end{equation}
The self-interacting terms of $\sigma$, $\om$ mesons and  the
isoscalar-isovector coupling  are given generally as
 \bey
 U_{\rm eff}(\sg,\om^\mu, b_0^\mu)&=&-\frac{1}{3}g_2\sg^3-\frac{1}{4}g_3\sg^4
 +\frac{1}{4}c_3(\om_\mu\om^\mu)^2\nnb\\
 &&+4\Ld_{V}g^2_\r  g_\om^2
 \om_\mu\om^\mu b_{0\mu}b_0^\mu. \label{eq:u}
 \eey
In addition, we include in the Lagrangian ${\cal L}_u$ for the U-boson
that is beyond the standard model. Following the form of the
vector meson, ${\cal L}_u$  is written as:
 \bey {\cal L}_u&=&-{\overline\psi}g_u\gm_{\mu}u^{\mu}\psi-\frac{1}{4}U_{\mu\nu} U^{\mu\nu}
      +\frac{1}{2}m_u^{2}u_{\mu}u^{\mu},
      \eey
with $u$ the field of U-boson. $U_{\mu\nu}$ is the strength tensor
of U-boson,
\begin{equation} U_{\mu\nu}=\pp_\mu u_{\nu} -\pp_\nu u_{\mu}.
\end{equation}

With the standard Euler-Lagrange formalism, we can deduce from the
Lagrangian the equations of motion for the nucleon and meson
fields. While in the mean-field approximation the Dirac field of nucleons
is quantized~\cite{Ser86},  the fields of mesons and U-boson,
which are replaced by their classical expectation values, obey
following equations: \bey
m_\sg^2\sg&=&g_\sg\r_s-g_2\sg^2-g_3\sg^3,\\
   m_\om^2\om_0&=&g_\om\r_B-c_3\om_0^3-8\Ld_{V} g_\r^2g_\om^2b_0^2\om_0,\\
   m_\r^2b_0&=&g_\r \r_3-8\Ld_{V} g_\r^2g_\om^2 \om_0^2b_0,\\
   m_u^2u_0&=&g_u\r_B,
\eey 
where the temporal subscript of the $\rho$ meson is neglected for convenience,  
$\r_s$ and $\r_B$ are the scalar and baryon densities,
respectively, and $\r_3$ is the difference between the proton and
neutron densities, namely, $\r_3=\r_p-\r_n$.  The set of coupled
equations can be solved self-consistently using the iteration
method. With these mean-field quantities, the resulting   energy
density $\vep$ and pressure $P$ for the hadronic phase are written as:
 \bey
 \vep&=&\sum_{i=p,n}\frac{2}{(2\pi)^3}\int^{k_{F_i}} d^3\!k E^*_i+
 \frac{1}{2}m_\om^2\om_0^2+\frac{1}{2}\frac{g_u^2}{m_u^2}\r_B^2+
 \frac{1}{2}m_\sg^2\sg^2+\frac{1}{2}m_\r^2 b_0^2 \nnb\\
&&+\frac{1}{3}g_2\sg^3+\frac{1}{4}g_3\sg^4+\frac{3}{4}c_3\om_0^4+12\Ld_{V}g^2_\r g_\om^2 \om_0^2 b_0^2,\label{eq:e}\\
P&=&\frac{1}{3}\sum_{i=p,n}\frac{2}{(2\pi)^3}\int^{k_{F_i}} d^3\!k
\frac{{\bf k}^2}{E^*_i}+\frac{1}{2}m_\om^2\om_0^2+\frac{1}{2}
\frac{g_u^2}{m_u^2}\r_B^2-\frac{1}{2}m_\sg^2\sg^2+\frac{1}{2}m_\r^2
b_0^2\nnb\\
&&-\frac{1}{3}g_2\sg^3-\frac{1}{4}g_3\sg^4+\frac{1}{4}c_3\om_0^4+4\Ld_{V}g^2_\r g_\om^2
 \om_0^2 b_0^2, \label{eq:p}\eey
with $E^*_i=\sqrt{{\bf k}^2+(M^*_i)^2}$.

For the quark phase, we use the MIT bag model,  in which the unique parameter, the bag
constant, arises from the energy difference between the
perturbative ground state and the  chiral symmetry
breaking vacuum of the theory~\cite{Fraga14}. In addition to its
simplicity, the MIT bag model has been widely used because of its
success in describing the vacuum properties of hadrons. As the
hadron-quark phase transition occurs, the hadronic degrees of
freedom starts to turn into quarks which are free of interactions
in the bag model. The pressure and energy density are given as:
 \bey
 \vep_Q&=&B+\sum_{f}\frac{3}{4\pi^2}[\mu_f(\mu_f^2-m_f^2)^{1/2}(\mu_f^2-\frac{1}{2}m_f^2)
 -\frac{1}{2}m_f^4\ln(\frac{\mu_f+(\mu_f^2-m_f^2)^{1/2}}{m_f})]\nnb\\&&+\frac{1}{2}
\frac{g_u^2}{m_u^2}\r_Q^2,\label{eq:eqp}\\
P_Q&=&-B+\sum_{f}\frac{1}{4\pi^2}[\mu_f(\mu_f^2-m_f^2)^{1/2}(\mu_f^2-\frac{5}{2}m_f^2)
 +\frac{3}{2}m_f^4\ln(\frac{\mu_f+(\mu_f^2-m_f^2)^{1/2}}{m_f})]\nnb\\&&+\frac{1}{2}
\frac{g_u^2}{m_u^2}\r_Q^2. \label{eq:pqp}\eey where $B$ is the bag
constant,  $\r_Q$ is the density of quarks, and the sum runs over
the favor $f$. With the perturbative QCD correction included to
the first order~\cite{Alford2005,Alcock86,Haensel86,Fraga01,xu15},
the grand thermodynamic potentials for quarks are given as
 \bey
 \Omega_u&=&-\frac{\mu_u^4}{4\pi^2}(1-c),\label{eq:ou1}\\
 \Omega_d&=&-\frac{\mu_d^4}{4\pi^2}(1-c),\label{eq:ou2}\\
   \Omega&=&\sum_{f=u,d,s}\Omega_f+B,
\eey where the term linear in $c={2\af_s}/{\pi}$ is from the
perturbative QCD correction, and the masses of up and down quarks
are set to zero in obtaining above expressions. Due to the nonzero
strange quark  mass ( $m_s=150 MeV$ in the calculation of this work), the
expression for the $\Omega_s$ is a little tedious and can be
referred to Ref.~\cite{Alcock86}. All thermodynamic quantities
follow consistently from the $\Omega$. Since quark matter in hybrid
stars is not in the perturbative regime, we regard  $c$ and the
bag constant as effective parameters in the present model, and
denote the effective perturbative QCD correction simply as the QCD correction
in the following.  The pressure is given by $P(\mu)=-\Omega(\mu)$,
the quark number density by $n(\mu)={\pp}P/{\pp}{\mu}$, and the
energy density by $\vep=-P+{\mu}n$.

We use Gibbs construction~\cite{Glendenning92,Glendenning01} to
depict the hadron-quark phase transition. After the hadron-quark
phase transition sets in, one may construct a mixed phase of
hadronic and quark matter over a finite range of pressures and
densities according to the Gibbs conditions for phase equilibrium.
The Gibbs conditions for the chemical and mechanical equilibriums
and the charge neutral condition are written as
 \bey
&&\mu_bb_i-\mu_cq_i=\mu_i,\\
&&P_H=P_Q,\\
&&(1-Y)\sum_{b}\r_{b}q_{b}+\frac{Y}{3}\sum_{f}\r_fq_f+\sum_{l}\r_lq_l=0,
 \eey
where $i$ runs over baryons, leptons  ($b_i=0$ for leptons) and
quarks, and $Y$ is the baryon number fraction of the quark phase.
With these conditions, the onset density of the hadron-quark phase
transition can be obtained, and the mixed phase is then
constructed. Eventually,  the total baryon density,  energy
density and pressure of the mixed phase are in turn given by
 \bey
&&(1-Y)\sum_{b}\r_{b}+\frac{Y}{3}\sum_{f}\r_f=\r,\\
&&\vep_M=(1-Y)\vep_H+Y\vep_Q+\vep_l,\\
&&P_M=(1-Y)P_H+YP_Q+P_l.
 \eey

Note that one may also use the Maxwell construction for the phase
transition. However, the Maxwell construction does not have the
mixed phase. In the Maxwell construction,  a direct transition
from hadronic to quark matter is accompanied by a density jump and
both phases  are separately charge neutral.
Regardless of the charge chemical equilibrium, only a single
(baryon) chemical potential is common to the hadronic and quark
phases with a constant pressure. To avoid the motion of charge due
to the different charge chemical potentials in two phases in the
Maxwell construction, a more realistic equation of state can be
obtained from the Wigner-Seitz cell calculation by taking into
account the Coulomb and surface effects. Then, the equation of state
of the mixed phase obtained from this approach becomes close to
that of the Gibbs construction~\cite{Yasutake14}.

Using the EOS of hybrid star matter as an input, we may obtain the
NS properties by  solving the Tolman-Oppenheimer-Volkoff (TOV)
equation~\cite{Op39,Tol39}:
\begin{eqnarray}
\frac{dP(r)}{dr}&=&-\frac{[P(r)+\vep(r)][M(r)+4\pi r^3 P(r)]}
{r(r-2M(r))},\label{eq:tov1}\\
M(r)&=&4\pi\int^r_0d\!\tilde{r}\tilde{r}^2
\vep(\tilde{r}),\label{eq:tov2}
\end{eqnarray}
where $r$ is the radial coordinate from the center of the star,
$P(r)$ and $\vep(r)$ are the pressure and energy density at the position
$r$, respectively, and $M(r)$ is the mass contained in the sphere of
the radius $r$. Note that here we use units for which the gravitation
constant is $G_\infty=c=1$. The radius $R$ and mass $M(R)$ of a
NS are obtained from the condition $p(R)=0$. Because the
NS matter undergoes a phase transition
from the homogeneous matter to the inhomogeneous matter at the low
density region, the EOS obtained from the homogeneous matter does
not apply to the low density region and the empirical low-density EOS in the
literature~\cite{Ba71,Ii97} is adopted.

\section{Results and discussions}
\label{results}

For the hadronic phase in hybrid stars,  we consider the simple
compositions:  neutrons, protons, electrons and muons. We do not
include hyperons in this work. The appearance of hyperons can
largely soften the equation of state, thus reducing the mass of
neutron stars greatly. Due to the accurate mass measurement of large-mass
neutron stars, people even conclude that the hyperon EOS has to be
ruled out~\cite{Demorest2010}. Indeed, the
onset densities of hyperons are very model-dependent. With the RMF
parameter set NL3, the $\Ld$ hyperon appears at about
$0.28fm^{-3}$ ~\cite{Jiang06,Yang08}, while it is at about $0.48fm^{-3}$
with the extended MDI interaction~\cite{Xu10}. The superfluidity of hyperons renders the $\Ld$
hyperons to appear at around $0.64fm^{-3}$~\cite{Takatsuka02},
which just affects the properties of neutron star slightly.
When the hyperonic degrees of freedom are included, it is found that the
hadron-quark transition density can not change continuously with
the emergence of hyperons. This means that the appearance of
hyperons  disfavors the hadron-quark phase transition in the RMF
models. Similar effect of the hyperons is found in the
literature~\cite{Chen11}. As implied by the large-mass NS's, 
the in-medium hyperon potential should be
density-dependent~\cite{jiang12}. We may further explore the
density-dependent interaction of hyperons in the future and it is
beyond the scope of the present work.

Among a number of nonlinear RMF models, we select two typical
best-fit parameter sets, NL3~\cite{La97} and FSUGold~\cite{Pie05},
to describe the hadron phase of hybrid star matter. Parameters and
saturation properties of these two parameter sets are listed in
Table~\ref{t:t1}. Usually, the nonlinear RMF models include the
nonlinear self-interacting meson terms  to simulate appropriate
in-medium effect of the strong interaction. The parameter set
NL3 includes the nonlinear self-interaction of the $\sigma$ meson,
while in addition to the latter,  the nonlinear self-interaction
of the $\omega$ meson is also included in  FSUGold. The resulting
EOS of the FSUGold is much softer than that of the NL3 at high
densities. It is known before that the vector U-boson can stiffen the
EOS, and  the stiffening role is much more significant for the
soft EOS in pure hadronic matter~\cite{zhang11}. It is thus
interesting to see whether we can observe a similar phenomenon in
hybrid star matter, if hadronic phase is described with these two
different RMF models.

\begin{table}[htb]
\caption{Parameters and saturation properties for the hadron phase
models NL3 and FSUGold. Meson masses, incompressibility $\kappa$
and symmetry energy are in units of MeV, and the density is in
unit of $fm^{-3}$. \label{t:t1}}
 \begin{center}
    \begin{tabular}{ c c c c c c c c c c c c c c c}
\hline\hline &$g_\sg$ & $g_\om$& $g_\r$& $m_\sg$ & $m_\om$  &
$m_\rho$ & $g_2$ & $g_3$& $c_3$ & $\Ld_V$ & $\rho_0$ &
$\kappa$ & $M^*/M$ & $E_{sym}$ \\
\hline
NL3     &10.217 &12.868 &4.474 &508.194 &782.501 &763 &10.431 &-28.890  &- &- &0.148 &271.8 &0.60 &37.4\\
FSUGold &10.592 &14.302 &5.884 &491.500 &782.500 &763 &4.277 &49.934 &418.39 &0.03 &0.148 &230.0 &0.61 &32.5\\
\hline\hline
\end{tabular}
\end{center}
\end{table}

While it is still an open question to determine the density at
which the hadron-quark phase transition occurs, we investigate the
onset of the phase transition in two manners: one is to fix the
transition density $\rho_c$ by adjusting the bag constant, and the
other to fix the bag constant. Given the bag
constant, we also obtain the transition density which is now
dependent on the hadron phase models. Once the transition density
is determined,  we can construct the mixed phase with the Gibbs
conditions and then obtain the EOS of hybrid star matter.

In this work, we choose the bag constant to be $B^{1/4}=180\sim220$
MeV which is a reasonable range between those values by fitting the
light-hadron
spectra~\cite{DeGrand75,Bartelski84,Chanowitz83,Carlson83} and those
(e.g., 250 MeV) used in the hydrodynamical evolution
of the quark gluon plasma~\cite{Ko89,He00}. The present range is
also close to that used in the
literature~\cite{Jiang13,Xu10,Weissenborn11}. For the choice of
the transition density,  we are referred to the literature where
it  is around $1.5\sim4\rho_0$
~\cite{Alford2005,Sharma07,Shao11,Cavagnoli11}. In addition to the
fact that the EOS of hybrid star matter with a
transition density $4\rho_0$ or even higher just has a minor
effect on the star maximum mass,  we choose the
range ($2\sim3\rho_0$) for the transition density in this
work.

As an example, we calculate here the various phase boundaries with
$B^{1/4}=180MeV$ and with the transition densities $2\rho_0$ and
$3\rho_0$.  For the parameter set NL3, the transition density with
$B^{1/4}=180MeV$ from the hadronic phase to the mixed phase is
about $0.20fm^{-3}$, and $0.77fm^{-3}$ from the mixed phase to
pure quark phase. The extent of the mixed phase is about $0.57
fm^{-3}$. For the parameter set FSUGold, the extent of the mixed
phase is $1.47fm^{-3}$ with the onset transition density
$0.28fm^{-3}$. For the case with the transition density fixed at
$2\rho_0$, the extent of the mixed phase is about $0.95fm^{-3}$
with the parameter set NL3, while it is $1.7fm^{-3}$ with the
parameter set FSUGold. For the transition density at $3\rho_0$,
the extents of the mixed phase are $1.99fm^{-3}$ and $2.34fm^{-3}$
with the NL3 and FSUGold, respectively.

\begin{figure}[thb]
\begin{center}
\vspace*{-5mm}
\includegraphics[height=8.0cm,width=10.0cm]{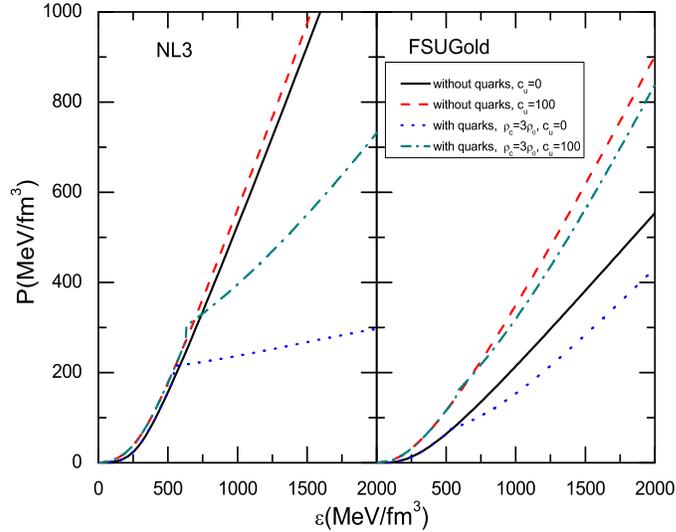}
 \end{center}
\vspace*{-5mm}\caption{(Color online) Equation of state of NS
matter with and without quarks. The curves with the U-boson
contribution are also shown for comparisons. The value of the
ratio parameter $c_u=(g_u/m_u)^2$ is in unit of $GeV^{-2}$. The
hadronic matter EOS is obtained with the NL3 (left panel)and
FSUGold (right panel). \label{eospt}}
\end{figure}

The phase transition usually tends to soften the EOS, as is
consistent with the requirement of  spontaneous stability in
natural processes. This is also true for the hadron-quark phase
transition in isospin asymmetric nuclear matter. In
Fig.~\ref{eospt}, we depict the EOS of hybrid star matter for
various cases with the onset density  $3\rho_0$ for hadron-quark
phase transition. In comparison to the EOS without phase
transition as shown in Fig.~\ref{eospt}, we see that the EOS of
isospin asymmetric matter is greatly softened by the hadron-quark
phase transition. It is striking to see that the stiff EOS with
the  NL3, which is not favored by the constraint from the flow
data of the heavy-ion reactions~\cite{da02},  even becomes much
softer than the soft FSUGold EOS due to the hadron-quark
transition.  With the inclusion of the U-boson, the soft EOS is
stiffened greatly.  Because the EOS of hybrid star matter
initiated with the NL3 is now  much softer than that with the
FSUGold, the stiffening effect turns out to be much more
appreciable for the EOS initiated with the NL3. The similar
stiffening role of the U-boson can also be clearly seen in the
case without phase transition, as we compare curves with the NL3
and FSUGold.   Since the softening of the EOS due to the
hadron-quark phase transition reduces largely the maximum mass of
NS's, we will see below that the inclusion of the U-boson plays an
important role in satisfying the maximum mass constraint for NS's.

\begin{figure}[thb]
\begin{center}
\vspace*{-5mm}
\includegraphics[height=12.0cm,width=12.0cm]{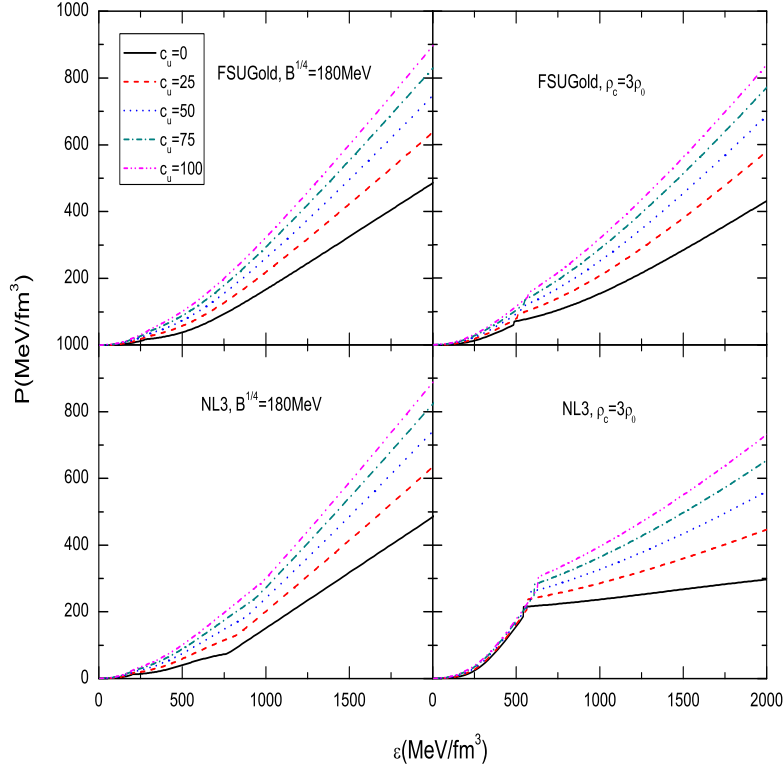}
 \end{center}
\vspace*{-5mm}\caption{(Color online) Equation of state of hybrid
star matter. The hadronic phase is given by two RMF parameter
sets, NL3 (lower panels) and FSUGold (upper panels).  For quark
phase, we present the results  with the fixed bag constant (left
panels ) and fixed transition density (right panels). In each
panel, the various curves are obtained with different ratio
parameters $c_u$. \label{eosnf}}
\end{figure}

To see the role of the U-boson specifically, we depict in
Fig.~\ref{eosnf} the EOS's of hybrid star matter  for a set of the
ratio parameter $c_u=(g_u/m_u)^2$.  Similar to that shown in
Fig.~\ref{eospt},  the EOS of hybrid star matter is stiffened
significantly  due to the repulsion provided by the U-boson. We
see that the EOS with the soft (FSUGold) and stiff (NL3) hadron
phase models acquires similar stiffening especially for the case
with the fixed bag constant. This is different from the case in
pure hadronic matter where a much more significant stiffening
effect is produced by the soft model~\cite{zhang11}. While for the
case with the fixed transition density, the stiffening role of
U-boson is more significant for the stiff parameter set NL3. The
reason for these to occur lies in the following facts. In pure
hadronic matter with RMF models, there is the cancelation between
the repulsion provided by the vector meson and the attraction provided
by the scalar meson. Thus, more significant cancelation in the
soft model sharpens the importance of the repulsion provided by
the U-boson.  While quarks are modeled by the MIT bag model with
the same bag constant, the  repulsion provided by the U-boson
plays the same stiffening role in the quark phase EOS, after the
hadron-quark phase transition occurs. In the case with the fixed
transition density, the quark phase EOS connected to the hadronic
phase with the NL3  is softer than that with the FSUGold, because
of the larger bag constant. Meanwhile, due to the phase
equilibrium in the mixed phase, the EOS initiated with the  NL3 is
softer than that with the FSUGold with increasing the density.
Thus, the U-boson provides a more significant stiffening role in
the EOS initiated with the stiff NL3 in the hadron phase.

\begin{figure}[thb]
\begin{center}
\vspace*{-5mm}
\includegraphics[height=9.0cm,width=8.0cm]{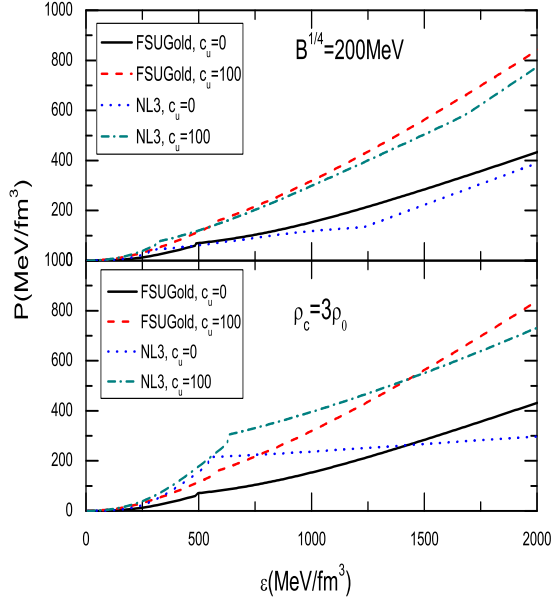}
 \end{center}
\vspace*{-5mm}\caption{(Color online) The influence of the U-boson
on the EOS of hybrid star matter for different hadron phase models
at the given bag constant $B=(200MeV)^4$ and transition density
$\r_c=3\r_0$, respectively. \label{MODE}}
\end{figure}

Since in pure hadronic matter the stiffening effect of the U-boson
is relevant to the extent of softness of the EOS's, it is
interesting to examine whether such a dependence exists in the EOS
of hybrid star matter with quark degrees of freedom. Shown in
Fig.~\ref{MODE} is the EOS's of hybrid star matter with and
without the contribution of the U-boson for two RMF parameter sets. We can
see that after the occurrence of the hadron-quark phase
transition, the difference in EOS's  with the fixed transition
density seems to be more apparent than that with  the fixed bag
constant. This is understandable, because the different bag
constants are used to obtain the same transition density. However,
compared to the large difference in the EOS's in pure hadronic
matter, as shown in Fig.~\ref{eospt}, we see that the phase
transition reduces  the difference largely in EOS's at high
densities. Compared  to the evolution of EOS's with the NL3 and
FSUGold, we see that the stiffer EOS undergoes a more appreciable
softening of the EOS in the mixed phase, as  the quark phase is
given by the same MIT bag model either with the fixed bag constant
or with a given transition density. We have also examined results
for the bag constants ranging from 180 MeV to 220 MeV and
transition densities from $3\r_0$ to $4\r_0$.  The increase of the
bag constant or transition density gives rise to a larger extent
of the mixed phase,  consistent with that found in
Ref.~\cite{prak97}. For instance, the transition density with
$B^{1/4}=200$ MeV is 0.28$fm^{-3}$ with the parameter set NL3, the
extent of the mixed phase increases from  0.57
$fm^{-3}$  (with $B^{1/4}=180$ MeV) to 1.1 $fm^{-3}$. With the
parameter set FSUGold, the extent of the mixed phase gets a more
apparent rise, and it is 2.8 $fm^{-3}$ starting from the
transition density 0.47 $fm^{-3}$. In any case, the appreciable
stiffening role of the U-boson in the soft EOS's can be  clearly
observed as in Fig.~\ref{MODE}, similar to that in
Fig.~\ref{eosnf}.

In this work, we make comparative study with the stiff and soft
EOS's. It is worthy to note the parameter $\Lambda_v$ may also
affect the EOS of asymmetric matter. The parameter $\Ld_v$ does
not appear in the parameter set NL3. In the parameter set FSUGold,
we may change this parameter (also change $g_\rho$) by keeping the
symmetry energy at saturation density unchanged.  Here, we choose
three parameter sets: FSUGw15 ($\Ld_v=0.015$), FSUGlod
($\Ld_v=0.03$), FSUGw45 ($\Ld_v=0.045$), all of which can fit the
ground-state properties of $^{208}$Pb~\cite{jia10a}. The
transition density increases  moderately with the rise of the
$\Ld_v$, since the stiffness of the symmetry energy that is
controlled by the $\Ld_v$ can affect the chemical equilibrium.
With the bag constant $B^{1/4}=180MeV$,  we obtain following
transition densities: 0.252, 0.282, and 0.297$fm^-3$ with the parameter sets
FSUGw15,  FSUGlod and FSUGw45, respectively.

\begin{figure}[thb]
\begin{center}
\vspace*{-5mm}
\includegraphics[height=12.0cm,width=12.0cm]{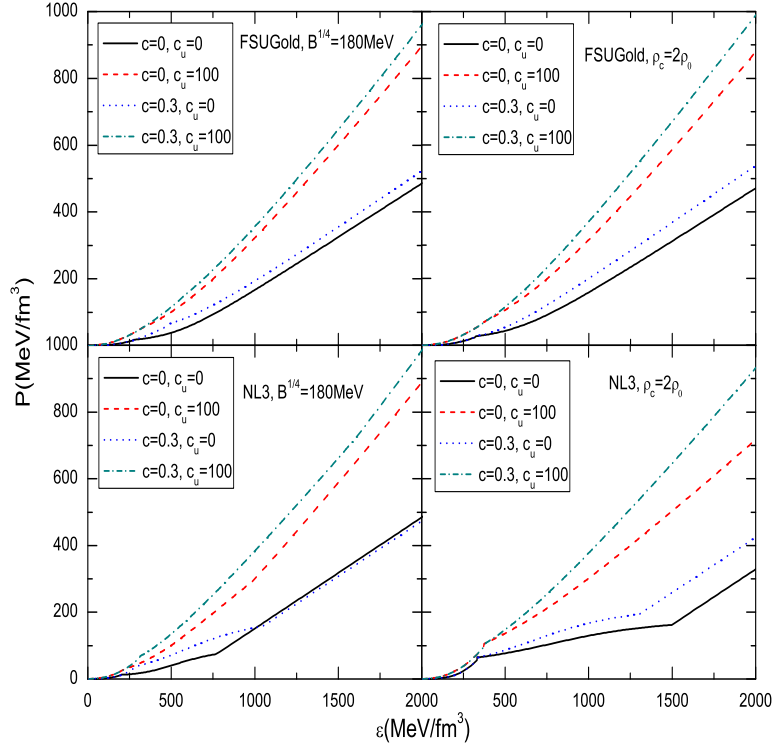}
 \end{center}
\vspace*{-5mm}\caption{(Color online) The EOS of hybrid star matter with the
inclusion of the QCD correction ($c=0.3$) with and without the U-boson.
\label{pqcd1}}
\end{figure}

In above, the quarks with the  bag model are free of interactions.
As the QCD coupling is not negligible  at densities of interest
for compact star physics, it is here necessary to discuss briefly
the QCD correction. It is reasonable to take the value of $c$ in
Eqs.(\ref{eq:ou1}),(\ref{eq:ou2}) to be 0.3~\cite{Alford2005}.
Shown in Fig.~\ref{pqcd1} is the EOS with the QCD correction. We
see from Fig.~\ref{pqcd1} that the QCD correction can stiffen
moderately the EOS of hybrid star matter at high densities.
The QCD correction may affect the phase transition
and increase the extent of the mixed phase due to the fact that the quark
phase pressure is now modified by the QCD correction.
For instance, with the bag constant $B^{1/4}=180MeV$, the onset
density of the mixed phase with the
parameter set NL3 is then about $0.28fm^{-3}$ with the extent of the mixed phase being about
$0.96fm^{-3}$. With the parameter set FSUGold, the onset density
of the mixed phase is about $0.46fm^{-3}$, and the extent of the
mixed phase grows dramatically up to $6.49fm^{-3}$.
As observed in Fig.~\ref{pqcd1}, the QCD correction to
the EOS is moderately dependent on the hadron phase models.
Specifically, the QCD correction results in an extent of the mixed
phase with the soft FSUGold  being larger than that with the stiff NL3.
We should say that the specific dependence of the  QCD correction
on the hadron phase models  is due to the connection built upon
the phase equilibrium conditions. For comparisons, we also plot
the results with and without the U-boson. As shown in
Fig.~\ref{pqcd1}, we see that the QCD correction is just moderate,
compared with the contribution of the U-boson. The stiffening role
of the U-boson is similar in cases with and without the QCD
correction.

\begin{figure}[thb]
\begin{center}
\vspace*{-5mm}
\includegraphics[height=12.0cm,width=12.0cm]{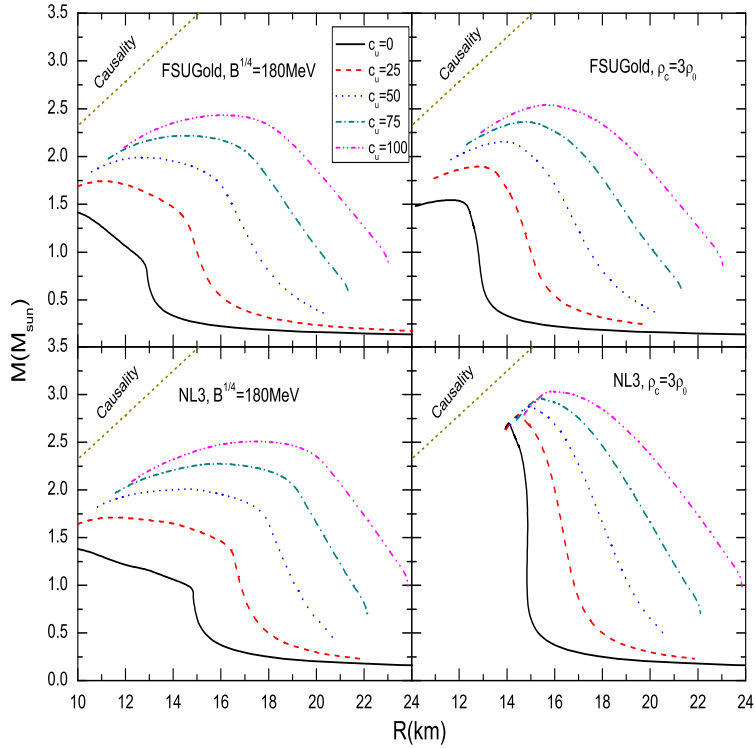}
 \end{center}
\vspace*{-5mm}\caption{(Color online) Mass-radius trajectories  of hybrid stars
with two hadron phase models. The U boson is included with  various
ratio parameters. \label{MR}}
\end{figure}

Now, we turn to the investigations of the hybrid star properties
with these EOS's discussed above. Shown in Fig.~\ref{MR} is the
relationship  between the NS mass and radius for various EOS's with
the inclusion of the U-boson. For the stiff hadron phase EOS, the
hadron-quark transition at a high density does not reduce the
maximum mass of hybrid stars significantly, as shown in the lower
right panel in Fig.~\ref{MR}. While in most cases, the phase
transition may results in very significant reduction of the
maximum mass, which should not be consistent with the observation
of large mass NS's~\cite{Demorest2010,Antoniadis2013}.
We see that the inclusion of the U boson can remedy the mass
eclipse very efficiently with the appropriately chosen ratio
parameter. The rise of the maximum mass is consistent with the
corresponding stiffening of the high-density EOS, shown in
Fig.~\ref{eosnf}.  We can see that the role of the U-boson in
increasing the maximum mass is more significant for softer EOS's.

Though shown in Fig.~\ref{MR} are only the results with the particular bag constant and transition density, it is sufficient to reveal the role of the U-boson since the change of these parameters yields rather similar results. It is, however, interesting to examine the consequences in the internal structure of the particular NS by varying these parameters. Let's first make comparison to the results with $B^{1/4}=180$ and 200 MeV. With
$B^{1/4}=180 MeV$, the parameter set NL3 gives a NS
maximum mass  $1.39 M_\odot$. The size of the quark core is 5.76
km, the mixed phase spreads from 5.76 to 7.62 km, and the hadronic
phase starts from 7.62 to 9.68 km. With $B^{1/4}=200 MeV$, the maximum mass is  $1.86 M_\odot$.
For convenient comparison, we also examine the internal structure of the  
$1.39 M_\odot$ NS with $B^{1/4}=200 MeV$.  The
size of the quark core is now 4.42 km, the mixed phase spreads
from 4.42 to 6.38 km, and the hadronic phase starts from 6.38 to
10.59 km. This indicates that the smaller bag constant that gives
rise to a smaller transition density features a larger quark core
in a particular NS. For the case
with the smaller transition density, the conclusion is similar
since the smaller transition density is given by the smaller bag
constant.

It is also interesting to examine the effect of the U-boson on the NS structure, 
although  the inclusion of the  U-boson  does not change the chemical and mechanical equilibriums,  the
transition densities, and the extents of each phases as well. As an example, here we examine the size of
the quark core and of the mixed phase for the maximum mass
configuration with the parameters $B^{1/4}=180 MeV$ and
$(g_u/m_u)^2=25GeV^{-2}$. With the inclusion of the
U-boson, the NS maximum mass  increases from $1.39
M_\odot$ to $1.71 M_\odot$. It is found in the NS
with the maximum mass configuration that the radius of
pure quark core is 5.3km, which is smaller than 5.76 km
obtained without the
inclusion of the U-boson. The shell of the mixed phase extends
from 5.3 to 8.6km, while without the inclusion of the U-boson, the
shell starts from 5.76 to 7.6 km. These results indicate that the
U-boson may change the ratio of various phases in the particular
NS though it does not take part in the phase
transition.

\begin{figure}[thb]
\begin{center}
\vspace*{-5mm}
\includegraphics[height=8.0cm,width=8.0cm]{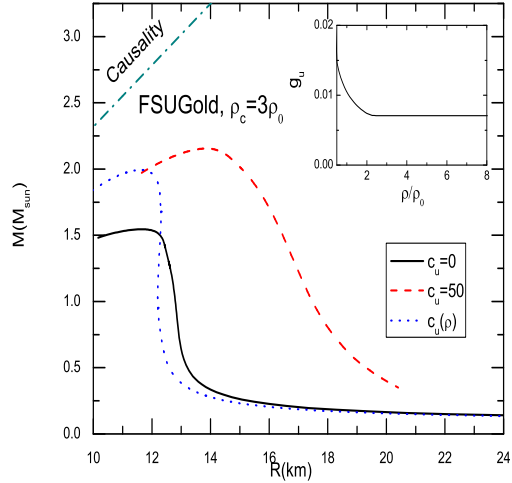}
 \end{center}
\vspace*{-5mm}\caption{(Color online) Mass-radius trajectories of
hybrid stars with the hybrid EOS initiated with the FSUGold. The
transition density is set to be $3\r_0$. The dotted curve  is the
result with the density-dependent coupling constant of the
U-boson, depicted in the inset. To obtain the density-dependent
coupling, the mass of the U-boson is taken as $1MeV$.
\label{gudd}}
\end{figure}
As shown in Fig.~\ref{MR}, the NS radius is significantly
increased by the U-boson. It was pointed out in the
literature~\cite{lat00,Li08} that the NS radius is primarily
determined by the EOS  in the lower density region of $1\rho_0$ to
$2\rho_0$. Since the inclusion of the U-boson also increases the
pressure in the lower density region appreciably, a large extent
of the NS radius is obtained accordingly with various $c_u$'s of
interest. This is similar to the previous works in
Ref.~\cite{Wen09,zhang11}.  It is known that the extraction of NS
radii from observations still suffers from large  systematic
uncertainties~\cite{mil13} involved in the distance measurements
and theoretical analyses of the light
spectrum~\cite{lat01,ha01,zh07,su11}. There is yet no consensus on
the extracted NS radii to date.   For instance, using the thermal
spectra from quiescent low-mass X-ray binaries (qLMXBs) Guillot
and collaborators extracted NS radii of $R_{\rm NS} = 9.4\pm1.2$
km~\cite{gu14}, while a relevant study of spectroscopic radius
measurements also suggest small radii $10.8^{+0.5}_{-0.4}$ km for
a 1.5 $M_\odot$ NS~\cite{oz15}.  There are also larger extracted
radii: a 3-$\sigma$ lower limit of 11.1 km on the radius of the
PSR J0437-4715~\cite{bog13} and a lower limit of 13 km for 4U
1608-52~\cite{pou14}. If the small radii of NS's is established,
we need to reconsider the large NS radii produced by the U-boson.
A possible solution of the NS radius suppression is to invoke the
appropriate in-medium effects~\cite{ji15}. Considering that the NS
radii are mainly determined by the low-density EOS, it is possible
to reduce the NS radii by constructing the density-dependent
coupling constant for the U-boson in the low-density regime. In
the high-density regime, we neglect the in-medium effect for the
U-boson, as one knows that the in-medium effect at high densities
is suppressed greatly by the Pauli blocking~\cite{br92,ma94,ji15}.
As an example, we perform the calculation with the EOS whose
hadron phase part is obtained with the FSUGold  and find that the
appropriate density-dependent coupling constant of the U-boson can
reduce the NS radii significantly, as shown in Fig.~\ref{gudd}.
The density-dependent coupling constant, depicted in the inset of
Fig.~\ref{gudd}, is designed to little change the pressure in the
low-density regime~\cite{ji15}. While the energy density still
acquires a significant increase from the U-boson, the resulting NS
radius is even less than the one obtained with the pure hadron
phase EOS. Since the NS maximum mass is determined mainly by the
high-density EOS, the present form of the  density dependence just
gives rise to a moderate reduction of the NS maximum mass.

\begin{figure}[thb]
\begin{center}
\vspace*{-5mm}
\includegraphics[height=8.0cm,width=9.0cm]{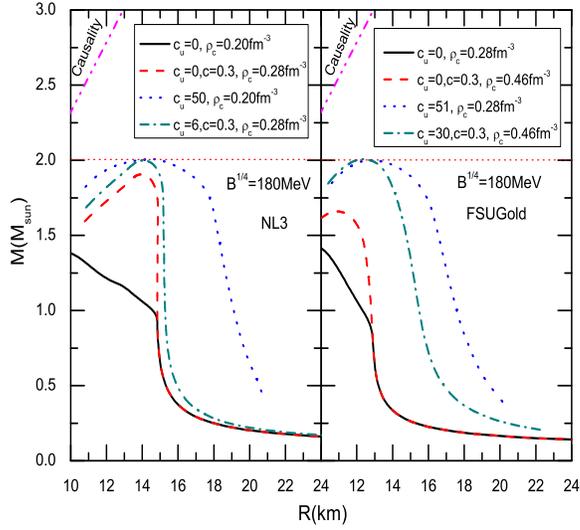}
 \end{center}
\vspace*{-5mm}\caption{(Color online) Mass-radius trajectories of
hybrid stars with and without  the QCD correction. To obtain the
2$M_\odot$ stars, the U-boson is also needed with the parameter
denoted in the legend. \label{fmr2}}
\end{figure}

In Figs.~\ref{MR} and \ref{gudd}, we have not included the
contribution of the QCD correction. The influence of the QCD
correction on the NS maximum mass is rather sensitive to the
transition  density. Given large transition densities, the QCD
correction just plays limited role in enhancing the NS maximum
mass, since the maximum mass in this case is dominated by the
hadron phase EOS. As the transition density decreases, the QCD
correction brings more significant enhancement for the NS maximum
mass. This findings are consistent with those in
Ref.~\cite{Weissenborn11}. On the other hand, as we fix the bag
constant, the situation for the QCD correction is different
because the transition density itself in this case is increased by
the correction through the phase equilibrium conditions.  As shown
in Fig.~\ref{fmr2}, the QCD correction with the fixed bag constant
gives rise to a significant enhancement of the NS maximum mass,
which is mainly attributed to the appreciable rise of the
transition density. In this case, since the hadron phase EOS with
the NL3 is much stiffer than that with the FSUGold, the large
transition density due to the QCD correction leads to a more
appreciable enhancement of the NS maximum mass with the NL3. With
the inclusion of the U-boson, the NS maximum mass can further
increase to satisfy the $2M_\odot$
constraint~\cite{Demorest2010,Antoniadis2013}. It is interesting
to see that we just need  a smaller ratio parameter of the U-boson
to meet the maximum mass constraint in this case. This is more
apparent for the EOS with the stiff NL3: the ratio parameter is
largely reduced by the QCD correction, as shown in Fig.~\ref{fmr2}

Since the magnitude and in-medium behavior of the U-boson ratio
parameter are important for the NS properties, it remains
significant to discuss the parameters for the U-boson. To satisfy
the NS maximum mass constraint, the ratio parameter $c_u$ in this
work is estimated to be around $0\sim 50GeV^{-2}$, depending on
the hadron phase models chosen. To avoid the violation of the
low-energy nuclear constraints for finite nuclei, we may limit the
weak interaction strength of the U-boson with the mass being of
order $1MeV$~\cite{zhang11,ji15}. For instance, if $g_u$ is 0.01,
then the mass of the U-boson would be below $1.4MeV$, responsible
for a long-range interaction.  The weak interaction strength of
the U-boson  does not compromise the success of nuclear models in
reproducing the properties of finite nuclei~\cite{xu13}.
Interestingly, the inclusion of the QCD correction may reduce the
ratio parameter significantly. We note that these estimated orders
of magnitude for the U-boson parameters can be compatible with
parameter regions allowed by a few experimental
constraints~\cite{Kr09}. Moreover, the density-dependent parameter
of the U-boson is found to be consistent with the usual
predictions on the NS radius, while the density dependence would
originate from the in-medium effect in the nuclear many-body
system~\cite{ji15,br92,ma94}.

\section{Summary}
\label{summary}

In this work, we have investigated the hadron-quark phase
transition using the Gibbs conditions with the RMF models for
hadron phase and MIT bag model for quark phase. We have considered
the U-boson to stiffen the EOS of hybrid matter that is softened
greatly by the phase transition.  With the inclusion of the
U-boson, the hybrid EOS is appreciably stiffened, and the
stiffening extent is similar in the EOS's with stiff and soft
hadron phase models  due to the fact that the same MIT model is
used for the quark phase. As a result, the NS maximum mass is
significantly increased by the U-boson.  In addition, we have
investigated the effect of the effective QCD correction on the
hybrid EOS. This  correction may give rise to the stiffening of
the hybrid EOS and the increase of the NS maximum mass that is
significant as the transition density is not very high. The
effective QCD correction can reduce the coupling strength of the
U-boson that is needed to  satisfy NS maximum mass constraint.
While the inclusion of the U-boson also increases the NS radius
significantly, we find that appropriate density dependence of the
U-boson coupling constant may bring the NS radius down to the
regime consistent with the observations and other model
predictions. We have also discussed that with the weak interaction
strength the inclusion of the U-boson does not compromise the
success of conventional nuclear models in reproducing properties
of finite nuclei. In summary, the inclusion of the U-boson can
favorably allow the existence of quark degrees of freedom in the
NS interior that is declined by the large mass NS observations. On
the other hand, the future coincident measurements and more
precise extraction of the mass and radius of neutron stars may
also bring the constraint for the U-boson that is yet to be
detected   in the terrestrial laboratory.

\section*{Acknowledgement}

The work was supported in part by the National Natural Science
Foundation of China under Grant No. 11275048 and the China Jiangsu
Provincial Natural Science Foundation under Grant No. BK20131286.

\end{document}